
\def\pmb#1{\setbox0=\hbox{#1}%
  \hbox{\kern-.025em\copy0\kern-\wd0
  \kern.05em\copy0\kern-\wd0
  \kern-0.025em\raise.0433em\box0} }

\catcode`@=11
\def\leftrightarrowfill{$\m@th\mathord\leftarrow \mkern-6mu
  \cleaders\hbox{$\mkern-2mu \mathord- \mkern-2mu$}\hfill
  \mkern-6mu \mathord\rightarrow$}
\def\overleftrightarrow#1{\vbox{\ialign{##\crcr
     \leftrightarrowfill\crcr\noalign{\kern-1pt\nointerlineskip}
     $\hfil\displaystyle{#1}\hfil$\crcr}}}
\catcode`@=12

\def\approxge{\hbox {\hfil\raise .4ex\hbox{$>$}\kern-.75 em
\lower .7ex\hbox{$\sim$}\hfil}}
\def\approxle{\hbox {\hfil\raise .4ex\hbox{$<$}\kern-.75 em
\lower .7ex\hbox{$\sim$}\hfil}}

\def \abstract#1 {\vskip 0.5truecm\sepline\vskip 0.5truecm
$$\vbox{\hsize=15truecm\noindent #1}$$}
\def \SISSA#1#2 {\vfil\vfil\centerline{Ref. S.I.S.S.A. #1 CM (#2)}}
\def \PACS#1 {\vfil\line{\hfil\hbox to 15truecm{PACS numbers: #1 \hfil}\hfil}}

\def \hfigure
     #1#2#3       {\midinsert \vskip #2 truecm $$\vbox{\hsize=14.5truecm
             \seven\baselineskip=10pt\noindent {\bcp \noindent Figure  #1}.
                   #3 } $$ \vskip -20pt \endinsert }

\def \hfiglin
     #1#2#3       {\midinsert \vskip #2 truecm $$\vbox{\hsize=14.5truecm
              \seven\baselineskip=10pt\noindent {\bcp \hfil\noindent
                   Figure  #1}. #3 \hfil} $$ \vskip -20pt \endinsert }

\def \vfigure
     #1#2#3#4     {\dimen0=\hsize \advance\dimen0 by -#3truecm
                   \midinsert \vbox to #2truecm{ \seven
                   \parshape=1 #3truecm \dimen0 \baselineskip=10pt \vfill
                   \noindent{\bcp Figure #1} \pretolerance=6500#4 \vfill }
                   \endinsert }

%
\def \ref
     #1#2         {\smallskip \item{[#1]}#2}
\def \sepline     {\medskip\centerline{\vbox{\hrule width5truecm}} \medskip}

\def \tabrul2     {\noalign{\vskip 5truept \hrule \vskip 2truept \hrule
                   \vskip 5truept} }


\footline={\ifnum\pageno>0 \tenrm \hss \folio \hss \fi }

\def\today
 {\count10=\year\advance\count10 by -1900 \number\day--\ifcase
  \month \or Jan\or Feb\or Mar\or Apr\or May\or Jun\or
             Jul\or Aug\or Sep\or Oct\or Nov\or Dec\fi--\number\count10}

\def\hour{\count10=\time\count11=\count10
\divide\count10 by 60 \count12=\count10
\multiply\count12 by 60 \advance\count11 by -\count12\count12=0
\number\count10 :\ifnum\count11 < 10 \number\count12\fi\number\count11}

\def\draft{
   \baselineskip=20pt
   \def\makeheadline{\vbox to 10pt{\vskip-22.5pt
   \line{\vbox to 8.5pt{}\the\headline}\vss}\nointerlineskip}
   \headline={\hfill \seven {\bcp Draft version}: today is \today\ at \hour
              \hfill}
          }

%
%

%
\catcode`@=11
%
%
\def\b@lank{ }

\newif\if@simboli
\newif\if@riferimenti

\newwrite\file@simboli
\def\simboli{
    \immediate\write16{ !!! Genera il file \jobname.SMB }
    \@simbolitrue\immediate\openout\file@simboli=\jobname.smb}

\newwrite\file@ausiliario
\def\riferimentifuturi{
    \immediate\write16{ !!! Genera il file \jobname.AUX }
    \@riferimentitrue\openin1 \jobname.aux
    \ifeof1\relax\else\closein1\relax\input\jobname.aux\fi
    \immediate\openout\file@ausiliario=\jobname.aux}

\newcount\eq@num\global\eq@num=0
\newcount\sect@num\global\sect@num=0

\newif\if@ndoppia
\def\numerazionedoppia{\@ndoppiatrue\gdef\la@sezionecorrente{\the\sect@num}}

\def\se@indefinito#1{\expandafter\ifx\csname#1\endcsname\relax}
\def\spo@glia#1>{} 

\newif\if@primasezione
\@primasezionetrue

\def\s@ection#1\par{\immediate
    \write16{#1}\if@primasezione\global\@primasezionefalse\else\goodbreak
    \vskip\spaziosoprasez\fi\noindent
    {\bf#1}\nobreak\vskip\spaziosottosez\nobreak\noindent}
%

\def\sezpreset#1{\global\sect@num=#1
    \immediate\write16{ !!! sez-preset = #1 }   }

\def\spaziosoprasez{50pt plus 60pt}
\def\spaziosottosez{15pt}

\def\sref#1{\se@indefinito{@s@#1}\immediate\write16{ ??? \string\sref{#1}
    non definita !!!}
    \expandafter\xdef\csname @s@#1\endcsname{??}\fi\csname @s@#1\endcsname}

\def\adv#1{\global\advance\sect@num by #1}

\def\autosez#1#2#3\par{
    \global\advance\sect@num by 1\if@ndoppia\global\eq@num=0\fi
    \xdef\la@sezionecorrente{\the\sect@num}
    \def\usa@getta{1}\se@indefinito{@s@#1}\def\usa@getta{2}\fi
    \expandafter\ifx\csname @s@#1\endcsname\la@sezionecorrente\def
    \usa@getta{2}\fi
    \ifodd\usa@getta\immediate\write16
      { ??? possibili riferimenti errati a \string\sref{#1} !!!}\fi
    \expandafter\xdef\csname @s@#1\endcsname{\la@sezionecorrente}
    \immediate\write16{\la@sezionecorrente. #2}
    \if@simboli
      \immediate\write\file@simboli{ }\immediate\write\file@simboli{ }
      \immediate\write\file@simboli{  Sezione
                                  \la@sezionecorrente :   sref.   #1}
      \immediate\write\file@simboli{ } \fi
    \if@riferimenti
      \immediate\write\file@ausiliario{\string\expandafter\string\edef
      \string\csname\b@lank @s@#1\string\endcsname{\la@sezionecorrente}}\fi
    \goodbreak\vskip 48pt plus 60pt
\centerline{\lltitle #2}                     
\centerline{\lltitle #3}                     
\par\nobreak\vskip 15pt \nobreak\noindent}

\def\semiautosez#1#2\par{
    \gdef\la@sezionecorrente{#1}\if@ndoppia\global\eq@num=0\fi
    \if@simboli
      \immediate\write\file@simboli{ }\immediate\write\file@simboli{ }
      \immediate\write\file@simboli{  Sezione ** : sref.
          \expandafter\spo@glia\meaning\la@sezionecorrente}
      \immediate\write\file@simboli{ }\fi
\noindent\lltitle \s@ection#2 \par}


\def\eqpreset#1{\global\eq@num=#1
     \immediate\write16{ !!! eq-preset = #1 }     }

\def\eqref#1{\se@indefinito{@eq@#1}
    \immediate\write16{ ??? \string\eqref{#1} non definita !!!}
    \expandafter\xdef\csname @eq@#1\endcsname{??}
    \fi\csname @eq@#1\endcsname}

\def\eqlabel#1{\global\advance\eq@num by 1
    \if@ndoppia\xdef\il@numero{\la@sezionecorrente.\the\eq@num}
       \else\xdef\il@numero{\the\eq@num}\fi
    \def\usa@getta{1}\se@indefinito{@eq@#1}\def\usa@getta{2}\fi
    \expandafter\ifx\csname @eq@#1\endcsname\il@numero\def\usa@getta{2}\fi
    \ifodd\usa@getta\immediate\write16
       { ??? possibili riferimenti errati a \string\eqref{#1} !!!}\fi
    \expandafter\xdef\csname @eq@#1\endcsname{\il@numero}
    \if@ndoppia
       \def\usa@getta{\expandafter\spo@glia\meaning
       \la@sezionecorrente.\the\eq@num}
       \else\def\usa@getta{\the\eq@num}\fi
    \if@simboli
       \immediate\write\file@simboli{  Equazione
            \usa@getta :  eqref.   #1}\fi
    \if@riferimenti
       \immediate\write\file@ausiliario{\string\expandafter\string\edef
       \string\csname\b@lank @eq@#1\string\endcsname{\usa@getta}}\fi}

\def\autoeqno#1{\eqlabel{#1}\eqno(\csname @eq@#1\endcsname)}
\def\autoleqno#1{\eqlabel{#1}\leqno(\csname @eq@#1\endcsname)}
\def\eqrefp#1{(\eqref{#1})}


\def\eq{\autoeqno}
\def\req{\eqrefp}
\def\chap{\autosez}        



\newcount\cit@num\global\cit@num=0

\newwrite\file@bibliografia
\newif\if@bibliografia
\@bibliografiafalse

\def\lp@cite{[}
\def\rp@cite{]}
\def\trap@cite#1{\lp@cite #1\rp@cite}
\def\lp@bibl{[}
\def\rp@bibl{]}
\def\trap@bibl#1{\lp@bibl #1\rp@bibl}

\def\refe@renza#1{\if@bibliografia\immediate        
    \write\file@bibliografia{
    \string\item{\trap@bibl{\cref{#1}}}\string
    \bibl@ref{#1}\string\bibl@skip}\fi}

\def\ref@ridefinita#1{\if@bibliografia\immediate\write\file@bibliografia{
    \string\item{?? \trap@bibl{\cref{#1}}} ??? tentativo di ridefinire la
      citazione #1 !!! \string\bibl@skip}\fi}

\def\bibl@ref#1{\se@indefinito{@ref@#1}\immediate
    \write16{ ??? biblitem #1 indefinito !!!}\expandafter\xdef
    \csname @ref@#1\endcsname{ ??}\fi\csname @ref@#1\endcsname}

\def\c@label#1{\global\advance\cit@num by 1\xdef            
   \la@citazione{\the\cit@num}\expandafter
   \xdef\csname @c@#1\endcsname{\la@citazione}}

\def\bibl@skip{\vskip +4truept}


\def\stileincite#1#2{\global\def\lp@cite{#1}\global   
    \def\rp@cite{#2}}                                 
\def\stileinbibl#1#2{\global\def\lp@bibl{#1}\global   
    \def\rp@bibl{#2}}                                 

\def\citpreset#1{\global\cit@num=#1
    \immediate\write16{ !!! cit-preset = #1 }    }

\def\autobibliografia{\global\@bibliografiatrue\immediate
    \write16{ !!! Genera il file \jobname.BIB}\immediate
    \openout\file@bibliografia=\jobname.bib}

\def\cref#1{\se@indefinito                  
   {@c@#1}\c@label{#1}\refe@renza{#1}\fi\csname @c@#1\endcsname}

\def\cite#1{\trap@cite{\cref{#1}}}                  
\def\ccite#1#2{\trap@cite{\cref{#1},\cref{#2}}}     
\def\ncite#1#2{\trap@cite{\cref{#1}--\cref{#2}}}    
\def\upcite#1{$^{\,\trap@cite{\cref{#1}}}$}               
\def\upccite#1#2{$^{\,\trap@cite{\cref{#1},\cref{#2}}}$}  
\def\upncite#1#2{$^{\,\trap@cite{\cref{#1}-\cref{#2}}}$}  

\def\clabel#1{\se@indefinito{@c@#1}\c@label           
    {#1}\refe@renza{#1}\else\c@label{#1}\ref@ridefinita{#1}\fi}

\def\biblskip#1{\def\bibl@skip{\vskip #1}}           

\def\insertbibliografia{\if@bibliografia             
    \immediate\write\file@bibliografia{ }
    \immediate\closeout\file@bibliografia
    \catcode`@=11\input\jobname.bib\catcode`@=12\fi}


\def\commento#1{\relax}
\def\biblitem#1#2\par{\expandafter\xdef\csname @ref@#1\endcsname{#2}}


\catcode`@=12


\magnification=1200
\topskip 20pt
\def\interlinea{\baselineskip=16pt}
\def\standardpage{\vsize=21.0truecm\voffset=+0.8truecm
                  \hsize=15.truecm\hoffset=+10truemm
                  \parindent=1.2truecm}

\tolerance 100000
\biblskip{+8truept}                        
\def\hbup{\hfill\break\baselineskip 16pt}  


\global\newcount\notenumber \global\notenumber=0
\def\note #1 {\global\advance\notenumber by1 \baselineskip 10pt
              \footnote{$^{\the\notenumber}$}{\nine #1} \interlinea}



\font\text=cmr10

\font\it=cmti10

\font\title=cmbx10 scaled \magstep3      
\font\ltitle=cmbx12 scaled \magstep1
\font\lltitle=cmbx12

\font\abs=cmti10 scaled \magstep1        

\font\seven=cmr7                         
\font\nine=cmr9                         
\font\bcp=cmbx7








\def\gtrsim{\ \rlap{\raise 2pt \hbox{$>$}}{\lower 2pt \hbox{$\sim$}}\ }
\def\lesssim{\ \rlap{\raise 2pt \hbox{$<$}}{\lower 2pt \hbox{$\sim$}}\ }


\def\mn{\medskip\noindent}
\def\bs{\bigskip}
\def\hb{\hfil\break}

\def\scss{\scriptscriptstyle}

\def\o{\over}



\def\ea{{\it et.al.}}
\def\ib{{\it ibid.\ }}

\def\npb#1{Nucl. Phys. {\bf B#1},}
\def\plb#1{Phys. Lett. {\bf B#1},}
\def\prd#1{Phys. Rev. {\bf D#1},}

\def\zpc#1{Z. Phys. {\bf C#1},}
\def\prep#1{Phys. Rep. {\bf #1},}


\stileincite{}{}     
\numerazionedoppia   

\interlinea
\standardpage
\text                


\def\scss#1{{\scriptscriptstyle #1}}

\def\o{\over}



\def\G{{\cal G_{\rm SM}}}
\def\E{{\rm E}_6}


\def\pr{\prime}


\font\mbf=cmmib10  scaled \magstep 2      

\def\bfe{{\hbox{\mbf\char'145}}}



\def\ra{\rangle}
\def\la{\langle}

\def\scss{\scriptscriptstyle }

\def\o{\over}









\def\E#1{{$E_{#1}\,$}}














\def\G{{\cal G_{\rm SM}}}
\def\E{{\rm E}_6}

\def\pr{\prime}


\font\mbf=cmmib10  scaled \magstep1      

\def\bfe{{\hbox{\mbf\char'145}}}


\autobibliografia
\pageno=0
\vsize=23.8truecm
\hsize=15.7truecm
\voffset=-1.truecm
\hoffset=+6truemm
\baselineskip 12pt
\rightline{SLAC-PUB--6422}\par\noindent
\rightline{UM-TH 93--29}\par\noindent
\rightline{January 1994}\par\noindent
\rightline{T/E}\par\noindent
\rightline{hep-ph/9401260}\par\noindent

\bs\bs\bs
\centerline{\ltitle
Identifying Unconventional E$_{\bf 6}$ Models}
\bs
\centerline{\ltitle at \bfe$^{\displaystyle\bf +}$
\bfe$^{\displaystyle\bf -}$ Colliders$^*$.}
\medskip
\vskip 1truecm                             

\centerline{Enrico Nardi$^1$ and Thomas G. Rizzo$^2$}
\bs
\bs
\centerline{\it $^1$Randall Laboratory of Physics, University of Michigan,
           Ann Arbor, MI 48109--1120}
\medskip
\centerline{\it $^2$Stanford Linear Accelerator Laboratory,
Stanford University, Stanford, CA 94309}

\vskip 1truecm                             
\medskip
\centerline{\bf {\abs Abstract}}           
\bs

\noindent
Recently it was shown that, in the framework of
superstring inspired $\E$ models, the presence of
generation dependent discrete symmetries allows us to construct
a phenomenologically viable class of models in which the three
generations of fermions do not have the same embedding within the
fundamental {\bf 27} dimensional representation of E$_6$.
In this scenario, these different embeddings of the conventional
fermions imply that the left-handed charged leptons and the
right-handed $d$-type quarks are coupled in a
non--universal way to the new neutral gauge bosons $(Z_\theta)$
present in these models.
It was also shown that a unique signature for this scenario,
would be a deviation from unity for the ratio of cross sections for the
production of two different lepton species in $e^+e^-$ annihilation.
However, several different scenarios are possible, depending on the particular
assignment chosen for $e_L$, $\mu_L$ and $\tau_L$ and
for the right-handed $d$-type quarks,
as well as on the type of $Z_\theta$ boson. Such scenarios can not be
disentangled from one another by means of cross section measurements alone.
In this paper we examine the possibility of identifying
the pattern of embeddings through measurements of
polarized and unpolarized asymmetries for fermion pair-production at the 500
GeV $e^+e^-$ Next Linear Collider (NLC).
We show that it will be possible to identify the different patterns of
unconventional assignments
for the left-handed leptons and for the $b_R$ quark, for
$Z_\theta$ masses as large as $\sim 1.5$ TeV.
\bs
\noindent
PACS number(s): 12.10.Dm,12.15.Ff,13.15.Jr,14.60.-z
\vfill
\noindent
--------------------------------------------\phantom{-} \hb
\leftline{Electronic addresses: nardi@umiphys.bitnet,
rizzo@scs.slac.stanford.edu}
\bigskip
\leftline{$^*$Work supported by the Department of Energy contract
DE-AC03-76SF000515}
\leftline{ and DE-AC02-76ER01112.}
                   \medskip

\eject

\standardpage                            
\interlinea                              


\null
\vskip -.5truecm
\chap{Int} {I. INTRODUCTION}{}

\noindent
Extended electroweak models based on the group $E_6$ are known to
lead to interesting phenomenology.
During the last few years,
a thorough analysis of these models was carried out
both from the theoretical and
phenomenological point of view [\cite{rizzo-e6}].
In particular, all the possible embeddings of the Standard Model
(SM) gauge group in $E_6$, as well as various possibilities for
the assignments of the matter fields to the fundamental representation
of the group, were analyzed in ref. [\cite{e6-emb}].

However, it has been recently shown [\cite{f}] that in the framework
of superstring derived E$_{\bf 6}$ models it is possible to implement
an unconventional scenario that has not been addressed by the previous
investigations. In this scenario some of the known fermions of the
three families are embedded in the fundamental {\bf 27} dimensional
representation of the group in a generation dependent way, implying
the possibility that corresponding fermions belonging to different
generations could have different gauge interactions with
respect to some subgroup of $\E$.

The realization of models with Unconventional Assignments (UA) for the
matter fields relies on the presence of generation dependent discrete
symmetries [\cite{f}]. Symmetries of this kind arise naturally
in field theories derived from the superstring, and for this reason
it can be argued
[\cite{f}] that the UA scenario should be considered as a natural
alternative to the standard schemes.

Of course, experimentally  we know that the $SU(2)\times U(1)$ gauge
interactions of the known fermions do respect universality with a high
degree of precision. However, since the SM gauge group is rank 4
while  $\E$ is rank  6, as many as two additional massive neutral
gauge bosons $(Z_\theta)$ can be present, possibly with $M_\theta \sim
1\>$TeV or less, and the possibility that
the corresponding new neutral interactions could violate universality is
still phenomenologically viable.

The 27 dimensional representation of $\E$ contains,
in addition to the standard model fermions,
two new leptonic $SU(2)$-doublets, two $SU(2)$-singlet
neutral states and two color-triplet $SU(2)$-singlet $d$-type quarks.
The UA models are realised by identifying
in a generation dependent way some of the
known doublets of left-handed (L) leptons
and/or singlets of right-handed (R) $d$-type quarks
with the new multiplets having the correct
$SU(2)\times U(1)$ transformation properties, in order to
ensure that the standard interactions
are unmodified with respect to the SM [\cite{f}].

For example, in the model described in Ref. [\cite{f}] the
L-handed lepton doublet ``${\nu_\tau \choose \tau}_L$'' and the
R-handed quark singlet ``$b_R$'' of the third generation are
assigned to the additional $SU(2)$ multiplets present in the {\bf 27}.
This model was shown to be consistent with a
large number of experimental constraints, ranging from the direct and
cosmological limits on the neutrino masses, to the stringent limits on
flavor changing neutral currents.
Clearly, since the different $SU(2)$ multiplets carry different $U_\theta(1)$
quantum numbers, in this model the ``$\tau_L$'' lepton and the ``$b_R$'' quark
have  a different interaction with the new $Z_\theta$
with respect to the corresponding states of the first two generations.

A very clean signature for the UA models would then be the detection
of deviations from universality induced by $Z_\theta$ exchange
in neutral current (NC) processes.
If UA are present in the leptonic sector
the experimentally clearest signature would be
a deviation from unity for the ratio of
production cross sections for two different lepton species
[\cite{fphen}].
A first study of the discovery limits for
lepton universality violation of this type in $e^+e^-$
annihilation at
the Large Electron Positron collider
with 180 GeV c.m. energy (LEP-2)
and at a 500
GeV $e^+e^-$ Next Linear Collider (NLC) [\cite{fphen}]
showed that such a signature could be detected
for a $Z_\theta$ mass as large as $\sim 800$
GeV (LEP-2) and $\sim 2$ TeV (NLC) [\cite{fphen}].

In fact, several different scenarios are possible,
depending on the particular assignment chosen for the
$e$, $\mu$ and $\tau$ leptons, and on the type of $Z_\theta$ boson,
and also UA could be present in the R-handed $d$-type quark sector.
While it is relatively easy to detect
possible deviations from universality induced by the UA,
it would be impossible to disentangle the different scenarios
by comparing the total cross sections alone.

In this paper we examine the possibility of identifying
the pattern of embeddings for the leptons and for the $b$ quarks
through measurements of polarized and unpolarized
asymmetries for the process $e^+e^-\to f\bar f$.
In particular we will analyse the discovery potential of the NLC assuming a
center of mass energy of 500 GeV.

In Sec. 2 we will briefly outline the main features of the $\E$ models
based on the UA scenario, and establish our conventions
and notations. A more complete discussion of the theoretical framework
can be found in Ref. [\cite{f}].

In Sec. 3 we will investigate the phenomenology of UA models
at future $e^+e^-$ machines.
We will give the relevant expressions for the
various polarized and unpolarized asymmetries and we will discuss
our results.
We stress that the large amount of data collected
at the $Z_0$ resonance by the LEP collaborations
are not effective in probing for these kind of effects.
In fact, as we have already mentioned,
the UA for the known fermions would not
affect the couplings to the standard $Z_0$,
while the contributions of $Z_\theta$--$Z_0$ interference
and of pure $Z_\theta$ exchange
to the various cross sections and asymmetries measured at LEP-1
are too small to be measured at the peak.
Some effects could still be detected, however,
if the $Z_0$ had a sizeable mixing with the $Z_\theta$
so that the universality of the couplings of the $Z_0$ to the usual fermions
would be indirectly affected by this mixing.
However, the existing bounds  on the  $Z_0$--$Z_\theta$
mixing angle are relatively tight [\cite{zp-new}] so that this possibility
seems unlikely, and  we will disregard it throughout this paper.
Finally in Sec. 4 we will summarize our results and draw our
conclusions.

We note that other authors have considered the possibility of using the NLC
to explore $Z'$ couplings for masses in excess of the collider's center of
mass energy[\cite{oldnlc1}][\cite{oldnlc2}][\cite{newnlc}] within a
more general context. It this
work we will focus on the specific problem of the ambiguity in the generation
dependent fermion quantum number assignments which can arise in $E_6$ models.

\vfill\eject
\chap{flip} {II. UNCONVENTIONAL ASSIGNMENTS}{IN E$_{\bf 6}$ MODELS}

In $\E$ grand unified theories,
as many as two new neutral gauge bosons
can be present, corresponding to the two additional
Cartan generators that are not present in the SM gauge group.
Here we will consider the embedding of the SM gauge group
$$
\G\equiv SU(3)_c\times SU(2)_L\times U(1)_Y  \eq{2.1}
$$
in $\E$
through the maximal subalgebras chain:

\setbox3=\hbox{\phantom{+}\raise6pt\hbox{\big\vert\kern-.7mm\lower6pt\hbox
{$\longrightarrow \ U(1)_{\chi} \times SU(5)$}}}
\setbox4=\hbox{\phantom{+}\raise6pt\hbox{\big\vert\kern-.7mm\lower6pt\hbox
{$\longrightarrow \ \G $.}}}

$$
\eqalign{\E \ \longrightarrow \ U(1)_{\psi} \times SO&(10)  \cr
&\copy3 \cr
&\phantom{\longrightarrow \ U(1)_{\chi} \times SU(}
\copy4 \cr}                                  \eq{2.2}
$$
\medskip\noindent
In general the two additional neutral gauge boson
will correspond to some linear combinations
of the $U_\chi(1)$ and $U_\psi(1)$ generators that we will parametrize
in term of an angle $\theta$ as
$$
\eqalign{
&Z_\theta^\pr= Z_\psi \cos\theta - Z_\chi \sin\theta  \cr
&Z_\theta^{\pr\pr}= Z_\psi \sin\theta + Z_\chi \cos\theta, \cr}
\eq{2.3}
$$
\medskip\noindent
The angle $\theta$ is a model dependent parameter
whose value is determined by
the details of the breaking of the gauge symmetry.
In the following we will denote the
lightest of the two new gauge bosons as $Z_\theta$.

In the kind of models we are considering here, each generation of
matter fields belong to one fundamental {\bf 27}
representation of the group. The {\bf 27} branches to
the ${\bf 1} + {\bf 10}
+ {\bf 16}$ representations of $SO(10)$.
The known particles of the three generations,
together with an $SU(2)$ singlet neutrino ``$\nu^c$",  are usually
assigned to the {\bf 16} of $SO(10)$, that in turn branches to
${\bf 1}_{\bf 16}$ + ${\bf \bar 5}_{\bf 16}$  + ${\bf 10}_{\bf 16}$
of $SU(5)$.
Giving in parenthesis the Abelian charges
$Q_\psi$ and $Q_\chi$ for the different $SU(5)$ multiplets,
we have
\smallskip
$$
\eqalign{
&[{\bf 1}_{\bf 16}]\ (1c_\psi)\>(-5c_\chi)=
\big[\nu^c\big]                                          \cr
&[{\bf \bar 5}_{\bf 16}]\ (1c_\psi)\>(3c_\chi)=
\big[L={\nu \choose e}, \, d^c \big]                        \cr
&[{\bf 10}_{\bf 16}]\ (1c_\psi)\>(-1c_\chi)=
\big[Q = {u\choose d}, \, u^c, \, e^c \big]        \cr}
\eq{2.4}
$$
\medskip\noindent
The {\bf 10} of $SO(10)$ that branches to {\bf 5}$_{\bf 10}$ +
${\bf \bar 5}_{\bf 10}$  of $SU(5)$ contains the fields
$$
\eqalign{
&[{\bf 5}_{\bf 10}]\ (-2c_\psi)\>(-2c_\chi)=
\big[H = {{N\choose E}}, \,  h^c\big]     \cr
&[{\bf \bar 5}_{\bf 10}]\ (-2c_\psi)\>(2c_\chi)=
\big[H^c = {{E^c\choose N^c}}, \,h].       \cr}
\eq{2.5}
$$
Finally the singlet {\bf 1} of $SO(10)$ corresponds to
$$
[\>{\bf 1}_{\bf 1}\>]\ (4c_\psi)\>(0c_\chi) = [S^c].
\eq{2.6}
$$
According to the normalization
$\sum_{f=1}^{27}(Q_{\psi,\chi}^f)^2 =
\sum_{f=1}^{27}({1 \over 2}Y^f)^2=5$,
in \req{2.4}-\req{2.6}
we have respectively
$c_\psi = {1\o 6}\sqrt{5\o 2}$ and $c_\chi =
{1\o 6}\sqrt{2\o 3}$.
Matter fields will couple for example to the
$Z_\theta^\pr$ boson through the charge
$$
Q_\theta = Q_\psi \cos\theta - Q_\chi \sin\theta.  \eq{2.7}
$$
\noindent
As it is clear from the second lines in
\req{2.4} and \req{2.5},
there is an ambiguity in assigning the known states to the {\bf 27}
representation, since under the SM gauge group
the ${\bf \bar 5}_{{\bf 10}}$ in
the {\bf 10} of $SO(10)$ has the same field content as the ${\bf
\bar 5}_{{\bf 16}}$ in the {\bf 16}. The same ambiguity is
also present for the two $\G$ singlets, namely
 ${\bf 1}_{{\bf 1}}$ and ${\bf 1}_{{\bf 16}}$.
This ambiguity has no physical consequences as long as
the same assignments are chosen for {\it all} the
three generations of SM fermions, that we will collectively denote as
$\{\psi_{SM}\}$. In fact, it is easy
to verify that transforming the model parameter
$\theta$ according to
$$
\theta\to \theta^\prime= \tan^{-1}(\sqrt{15})-\theta
\eq{2.8}
$$
induces on the charges in \req{2.7}
the transformations
$Q_\theta({\bf \bar 5}_{\bf 16})
\to Q_\theta^\prime({\bf \bar 5}_{\bf 16}) =Q_\theta({\bf \bar 5}_{\bf 10})$
and $Q_\theta({\bf 1}_{\bf 16}) \to
Q_\theta^\prime({\bf 1}_{\bf 16}) = Q_\theta({\bf 1}_{\bf 10})$
while at the same time the charges for the
${\bf 5}_{\bf 10}$ and for the ${\bf 10}_{\bf 16}$
are left invariant.
This means that a model defined by a particular value of
$\theta$ and
by a choice of the assignments for the
three generations of  SM fermions
$\{ Z_\theta, \, \{\psi_{SM}\} \subset {\bf \bar 5}_{\bf 16}\}$
is physically equivalent to the model defined by
$\{ Z_{\theta^\prime}, \, \{\psi_{SM}\}
\subset {\bf \bar 5}_{\bf 10}\}$.

However, UA models realize a scenario in which the assignments
of the SM fermions to the two ${\bf \bar 5}$'s
are generation dependent.
For example in the model analyzed
in [\cite{f}] what we call ``$\tau_L$" corresponds
in fact
to the charged component of the $H_3$ weak doublet belonging to
${\bf \bar 5}_{{\bf 10}}$, while the
``$e_L$'' and the ``$\mu_L$'' leptons are  as usual assigned to the
${\bf \bar 5}_{{\bf 16}}$.
Since these fermions might not correspond to the entries as listed in
\req{2.4} we use quotation marks to denote the known states with
their  conventional labels, while labels not enclosed within quotation
marks will always refer to the $SU(2)$ multiplets in \req{2.3}--\req{2.5}.
This model was realized by imposing on the superpotential
a particular family-non-blind
$Z_2\times Z_3$ discrete symmetry.
As a result  of such a symmetry,
the masses of the known (light) chiral leptons
are generated by vacuum expectation values (VEVs) of Higgs doublets,
through the terms $m_\tau E_{3L}e_{3R}$
(with $m_\tau\sim \la\tilde L_3\ra_0$) and
$m_{\alpha\beta}e_{\alpha L}e_{\beta R}$
(with $m_{\alpha\beta} \sim \la\tilde H_2\ra_0$ and
$\alpha,\beta=1,2$).
The remaining charged leptons
$e_{3L}$, $E_{3R}$,
$E_{\alpha L}$, $E_{\alpha R}$ are vectorlike, and acquire large
masses from VEVs of Higgs singlets.

Clearly, also in the UA models of this kind it is not physically
meaningful to ask if the
L-doublet of leptons of the first generation sits in the
{\bf 16} or in the {\bf 10} of $SO(10)$ since these two choices
are equivalent under the transformation \req{2.8}.
However, experiments could tell us if for example the
L-doublet of leptons of the second generation
sits in the {\it same} $SO(10)$ multiplet or in a {\it different} one. That is,
experiment can tell us whether or not the transformation properties of the
three generations of fermions under the extended gauge group are the
same,
provided the scale associated with the breaking of this extended group is not
too high.

In the present paper we will concentrate on the phenomenological
consequences of
having different assignments for the standard L leptons
and for the ``$b_R$" quark.
We will present the theoretical predictions for several quantities
(cross sections and asymmetries)
measurable in the process $e^+e^-\to f\bar f$.
Referring to the initial and final states,
for each observable ${\cal A}$
we will label the two physically
different possibilities as ${\cal A}_{{\bf 16}\to {\bf 16}}$
and ${\cal A}_{{\bf 16}\to {\bf 10}}$.
However, according to the previous
discussion, it should be kept in mind that when the model parameter
$\theta$ is transformed according to \req{2.8} these observables are
found to directly
correspond to ${\cal A}_{{\bf 10}\to {\bf 10}}$ and
${\cal A}_{{\bf 10}\to {\bf 16}}$. Thus these latter observables are not
truly independent. That this is true can be seen most clearly in the case
of charm pair production since the charm couplings to the $Z_\theta$
are invariant
under \req{2.8} (as charm always lies in the ${\bf 16}$) and thus this
reaction can potentially only probe whether $e^-_L$ is placed in the
${\bf 16}$ or ${\bf 10}$. Fig. 1 shows the production cross section for charm
pairs at the NLC for a $Z_\theta$ mass of 1 TeV as a function of the parameter
$\theta$. The two curves show how the cross section apparently depends on the
choice of the $e^-_L$ assignment. However, we see under close examination that
the two curves are actually the {\it same} in that one can be obtained from
the other merely by making the transformation \req{2.8}. From this we learn
explicitly that ${\cal A}_{{\bf 10}\to {\bf 16}}$ and
${\cal A}_{{\bf 16}\to {\bf 16}}$ are simply related through \req{2.8}. The
same sort of relation can also be shown to exist between
${\cal A}_{{\bf 16}\to {\bf 10}}$ and ${\cal A}_{{\bf 10}\to {\bf 10}}$
through identical arguments.

\chap{phen} {III. IDENTIFYING THE UNCONVENTIONAL}
{ASSIGNMENTS AT \bfe$^{\displaystyle\bf +}$ \bfe$^{\displaystyle\bf -}$
COLLIDERS}

We will consider now the cross sections and the asymmetries
in the presence of additional neutral gauge bosons
for the process
$e^+(p_L^+)e^-(p_L^-)\to f\bar f$ ($f\neq e$),
where $p_L^\pm$ represent the longitudinal polarization of
$e^\pm$.
We will henceforth assume
that one of the two new bosons in \req{2.3} is  heavy enough so that
its effects on the low energy physics are negligible. Then
we will use the subscripts
$i,j=0$,$1$,$2$ which correspond respectively to the
$\gamma$, $Z_0$ and $Z_\theta$  bosons.
Denoting by
${\cal P}\equiv (p_L^+-p_L^-)/(1-p_L^+ p_L^-)$
the overall polarization of the initial
electron-positron system, the
cross section and the forward backward asymmetries for massless fermion pairs
can be written in terms of
$$
\eqlabel{3.1}
\eqlabel{3.2}
\eqlabel{3.3}
\eqlabel{3.4}
\eqalignno{
\sigma^f_{TOT}(s,p^+_L,p^-_L)&=
{4\o 3}{\pi\alpha^2\o s}\,(1-p_L^+p_L^-)
\,[T_1(s)+{\cal P}T_2(s)]
                                                  &\req{3.1}  \cr
\sigma^f_{FB}(s,p^+_L,p^-_L)&=
\>\>{\pi\alpha^2\o s}\>(1-p_L^+p_L^-)
\>[T_3(s)+{\cal P}T_4(s)]
                                                  &\req{3.2}  \cr
&&\cr
T_1(s)=\sum_{i,j=0}^2 &C_{ij}(e)\, C_{ij}(f)\,  \chi_i(s)\chi_j^*(s)
\cr            
T_2(s)=\sum_{i,j=0}^2 &C_{ij}(e)\, C^\prime_{ij}(f)\,  \chi_i(s)\chi_j^*(s)
\cr
T_3(s)=\sum_{i,j=0}^2 &C^\prime_{ij}(e)\, C^\prime_{ij}(f)\,
\chi_i(s)\chi_j^*(s)
\cr            
T_4(s)=\sum_{i,j=0}^2 &C^\prime_{ij}(e)\, C_{ij}(f)\chi_i(s)\chi_j^*(s)\,
\cr
C_{ij}=[v_iv_j &+ a_ia_j]
\cr                 
C_{ij}^\prime=[v_ia_j &+ a_iv_j]
&\req{3.3}   \cr
&&\cr
\chi_i(s)={g^2_i\o 4\pi\alpha}&{s\o s - M^2_i - iM_i\Gamma_i}.
                        &\req{3.4}
\cr }
$$
\mn
where the labels $e$ and $f$ in $T_1$,\dots $T_4$
refer to the couplings to the $Z_{i,j}$ bosons
of the electron and of the fermions in the final state.
The couplings in \req{3.3} and \req{3.4} are given by
$$
\eqlabel{3.5}
\vbox{
\baselineskip=18pt
\settabs \+
$g_1=(\sqrt{2}G_\mu M^2_Z)^{1\o 2}$ \quad
&$v_1(f)=T_{3L}^f-2Q_{\rm em}^f s^2_w$   \quad
&$a_2(f)=Q_\theta^f+Q_\theta^{f^c}$   \quad
&\quad$\ f=e,\mu,\tau$
\cr  \+
$g_0=e$
&$v_0(f)=Q_{\rm em}^f$
&$a_0(f)=0$
&
\cr  \+
$g_1=(\sqrt{2}G_\mu M^2_Z)^{1\o 2}$
&$v_1(f)=T_{3L}^f-2Q_{\rm em}^f s^2_w$
&$a_1(f)=T_{3L}^f$
&\phantom{\quad$\ f$}\req{3.5}
\cr  \+
$g_2=s_w g_1$
&$v_2(f)=Q_\theta^f-Q_\theta^{f^c}$
&$a_2(f)=Q_\theta^f+Q_\theta^{f^c}$
&
\cr  }
$$
\interlinea
\mn
where $Q_{\rm em}^f$ is the electric charge,
$T_{3L}^f$ is the third component of the weak isospin, and
$s_w\equiv \sin\theta_w$ with $\theta_w$ being
the weak mixing angle. For numerical purposes we take $\alpha(M_Z)^{-1}=127.9$,
$M_1=91.187$ GeV, $\Gamma_1=2.489$ GeV, and $\sin^2 \theta=0.2325$ in
evaluating the above expressions[\cite{leppho}]. In addition, for
simplicity we will also assume
that $\Gamma_2=0.01M_2$ for all values of the parameter $\theta$; our
results are insensitive to this assumption. The
couplings $Q_\theta^{f,f^c} =
Q_\psi^{f,f^c}\cos\theta - Q_\chi^{f,f^c}\sin\theta\ $
to the $Z_\theta$ boson have been defined in \req{2.7},
and the new charges
$Q_\psi^{f,f^c}$, $Q_\chi^{f,f^c}$ are given in parenthesis in
\req{2.4}-\req{2.6}.
In addition,
in \req{3.5} we have assumed
for the abelian coupling, $g_2$,
a renormalization
group evolution down to the electroweak scale
similar to that of
the hypercharge coupling $g_{\scss Y}\simeq s_w g_1$.

Since we are neglecting a possible
$Z_0$--$Z_\theta$ mixing,
the vector and axial-vector couplings,
$v_{0,1}(f)$ and $a_{0,1}(f)$ in \req{3.5}
do not  depend on the specific assignments, and
are unmodified with respect to the SM.
In contrast, $v_{2}(f)$ and $a_{2}(f)$
do depend on the particular assignments of the $f$ fermion
to the {\bf 16} or to the {\bf 10} representations
of $SO(10)$. Here we are interested in the cases
$f=e_L,\mu_L,\tau_L,b_R$.

The total cross section, the forward-backward, the polarized
forward-backward and the left-right asymmetries are defined in terms of
the above quantities by
$$
\eqlabel{3.6}
\eqlabel{3.7}
\eqlabel{3.8}
\eqlabel{3.9}
\eqalignno{
\sigma^f&=
{4\o 3}\>{\pi\alpha^2\o s}\>T_1                            &\req{3.6}
\cr
A^f_{FB}&={3\o 4}\>{T_3\o T_1}                         &\req{3.7}
\cr
A^{pol\> f}_{FB}&={}\>{3\o 4}\>{T_2\o T_1}     &\req{3.8}
\cr
A^f_{LR }&={}\>{T_4\o T_1}     &\req{3.9}
\cr}
$$
\mn
In the case of $c,b-$quark pair production, we include for numerical purposes
the leading order QCD corrections to the cross sections and asymmetries
taking $\alpha_s(M_Z)=0.123$ [\cite{leppho}] and employing the standard
three-loop renormalization group equations to control the running to the
500 GeV mass scale.

In order to give some feeling for the typical precision obtainable for the
measurement of the quantities defined above we will assume an integrated
luminosity of ${\cal L}=50 fb^{-1}$, which corresponds to approximately 2
years of
running at the NLC, and a single beam polarization of ${\cal P}=90\%$.
Furthermore, we will assume both these values are rather precisely determined:
$\delta {\cal L}/{\cal L}=0.6\%$ and $\delta {\cal P}/{\cal P}=1\%$;
in addition, we use a $\mu$- and
$\tau$-tagging efficiency of $100\%$ and a b-tagging efficiency of $80\%$.

Figs.~2 and 3 show the predicted values for the quantities \req{3.6}-\req{3.9}
as
functions of the parameter $\theta$ for lepton and $b$-quark pair production
respectively. Also shown in each case as a `data point' is the SM value and
error assuming the values above for ${\cal L}$, ${\cal P}$, etc. The
light(dark) shaded region in each case corresponds to the transition
${\bf 10} \to {\bf 16}({\bf 16} \to {\bf 16})$ with the `inner'(`outer')
edge of
each region corresponding to a $Z_{\theta}$ mass of 1.5 (1.0) TeV. The shaded
regions thus show not only the anticipated deviation from the predictions of
the SM but also how sensitive these are to the $Z_{\theta}$ mass.
(Presumably, by the time such experiments are performed at the NLC, the mass of
any new neutral gauge boson will have already been well determined at the LHC.)
Clearly if the $Z_{\theta}$ mass were much larger than 1.5 TeV it would be
quite difficult to distinguish between models without increased integrated
luminosity and a reduction of systematic errors. In Fig.~2, we see that the
${\bf 10} \to {\bf 16}$ and ${\bf 16} \to {\bf 16}$ cases are easily
separable for the 1 TeV possibility as long as the value of $\theta$ is
not too
close to $\sin^{-1}\sqrt{3\o 8}\simeq 37.76^{\rm o}$.
For this particular value the
$Z_{\theta}$ couplings are {\it independent} of the choice
${\bf 10} \to {\bf 16}$ or ${\bf 16} \to {\bf 16}$ since this value of
$\theta$ maps into itself under the transformation \req{2.8}. It is also clear
from these figures that the total cross-section, $\sigma^{l}$, and the
left-right polarization asymmetry, $A_{LR}^{l}$, play the dominant r\^ole in
identifying the model, while
for positive values of $\theta$
the forward-backward asymmetry $A_{FB}^{l}$  displays
a large overlap between the two regions
corresponding to the different assignments.
{}From this figure we can see that our results are in
general agreement, where they overlap, with those already existing in the
literature which have considered more general models
for $Z'$s interacting universally with the fermions
[\cite{oldnlc1}][\cite{oldnlc2}][\cite{newnlc}].

The situation is similar but a bit more difficult in
the $b$-quark case, as is shown in Fig.~3, the reason being the far greater
overlap of the ${\bf 10} \to {\bf 16}$ and ${\bf 16} \to {\bf 16}$ shaded
regions. Unlike the leptonic case, where the overlap between the two
sets of predictions occurs only near $37.76^{\rm o}$, most of the
quantities in the
$b$-quark case display 2 overlap regions, and generally lie somewhat closer to
the various SM
predictions. Here we see that in contrast to the leptonic case the quantity
$A_{FB}^{b}$ will play a very dominant r\^ole in distinguishing the two
scenarios. Combining all four observables allows the two embedding scenarios
to be separated for all $\theta$ values not too close to
$37.76^{\rm o}$,
provided that $Z_{\theta}$ is not far above 1.5 TeV.

Clearly, for large $Z_{\theta}$
masses, separating the various scenarios within $E_6$ becomes much more
difficult. From the figures, however, we see that the {\it relative}
assignments of the various leptons and the b-quark can be determined for
masses as large as 1.5 TeV.

\chap{sum} {IV. SUMMARY AND CONCLUSIONS}

\noindent
In $E_6$ models it is possible that the three generations may have different
embeddings into the fundamental {\bf 27} representation while still
maintaining universality as far as SM interactions are concerned. The various
embeddings can thus only be distinguished based upon probes of the model's
$Z_{\theta}$ couplings. Although we anticipate that such particles are more
massive than 500 GeV, the NLC allows us to indirectly probe the $Z_{\theta}$'s
couplings to leptons, $b$- and $c$-quarks thus allowing us
to distinguish between
various embedding scenarios. In order to do this we need large integrated
luminosities to reduce statistical uncertainties as well as good control of
experimental systematics.
The results of our analysis show that
for most of the value of the model dependent parameter $\theta$,
the various embedding schemes can be clearly separated
for $Z_\theta$ bosons as heavy as 1.5 TeV.
Only in a small region of the $\theta$ axis
around $\theta=37.76^{\rm o}$ the disentanglement
of the different embeddings is not possible, and this is
due to the fact that for this value of $\theta$ the different choices
are physically equivalent.
Clearly, use of both total cross section and the
various asymmetry measurements are needed to perform this analysis for leptons
and $b$-quarks simultaneously,
as no one piece of data alone is sufficient to separate
the various embedding schemes.

\chap{ackn} {V. ACKNOWLEDGEMENTS}

One of us (TGR) would like to thank J.L. Hewett for discussions related to
this work and the members of the Argonne National Laboratory Theory Group for
use of their computing facilities. This work was supported in part by the
Department of Energy, contracts DE-AC03-76SF000515 and
DE-AC02-76ER01112.

\vfill\eject

\null
\centerline{\title References}
\baselineskip 8pt

\vskip .8truecm

\biblitem{rizzo-e6}
See J.L. Hewett and T.G. Rizzo, \prep{183} 195 (1989) and references
therein. \par

\biblitem{e6-emb}
M. Dine \ea,  \npb{259} 549 (1985);\hbup
F. del Aguila, J.A. Gonzales and M. Quiros, \npb{307} 571 (1988). \par

\biblitem{f}
E. Nardi, \prd{48} 3277 (1993). \par

\biblitem{fphen}
E. Nardi, Report UM-TH-93-19, to appear on Phys. Rev. {\bf D}. \par

\biblitem{zp-new}
E. Nardi, E. Roulet and D. Tommasini, \prd{46} 3040 (1992); \hbup
P. Langacker and M. Luo, \ib {\bf 45} 278 (1992); \hbup
F. del Aguila, W. Hollik, J.M. Moreno and M. Quir\'os, \npb{372}
 3 (1992); \hbup
J. Layssac, F.M. Renard and C. Verzegnassi, \zpc{53} 97 (1992); \hbup
M.C. Gonzalez Garc\'\i a and J.W.F. Valle; \plb{259} 365 (1991); \hbup
G. Altarelli \ea, \ib {\bf 263} 459 (1991). \par

\biblitem{oldnlc1}
A. Djouadi, A. Leike, T. Riemann, D. Schaile and C. Verzegnassi in the
Proceedings of the {\it Workshop on Physics and Experiments with Linear
$e^+e^-$ Colliders}, September 1991, Saariselke\"a, Finland, R. Orava ed.,
Vol. II, p.515 and \zpc{56} (1992) 289. \par

\biblitem{oldnlc2}
J. Hewett and T.G. Rizzo in the
Proceedings of the {\it Workshop on Physics and Experiments with Linear
$e^+e^-$ Colliders}, September 1991, Saariselke\"a, Finland, R. Orava ed.,
Vol. II, pp. 489 and 501. \par

\biblitem{newnlc}
A. Leike, DESY report DESY 91-154, 1993; F. Del Aguila and M. Cvetic,
University of Pennsylvania report UPR-590-T, 1993. \par

\biblitem{leppho}
See talks given by W. Hollik and G. Coignet at the {\it XVI International
Symposium on Lepton-Photon Interactions}, Cornell University,
August 1993. \par

\insertbibliografia

\vfill\eject

\centerline{\lltitle Figure captions}
\interlinea

\bs
\noindent
FIG. 1.

\noindent
The total charm pair production cross section at the NLC as a function of the
$E_6$ parameter $\theta$ for the cases ${{\bf 16}\to {\bf 16}}$ (dashes) and
${{\bf 16}\to {\bf 10}}$ (dash-dots). The picture shows that
the curves overlap when one of them is
reflected with respect to the point
$\theta=37.76^{\rm o}$. A $Z_{\theta}$ with a mass of 1 TeV has been assumed.

\vskip 1.5truecm

\noindent
FIG. 2.

\noindent
Predicted values of the leptonic observables at the NLC as functions of
the $E_6$ parameter $\theta$. In
each case the data point represents the SM prediction and anticipated error.
The ${{\bf 16}\to {\bf 16}}$ case is represented by the heavier shading while
the ${{\bf 16}\to {\bf 10}}$ case has a lighter shading. The inner (outer)
boundaries in all cases correspond to a $Z_{\theta}$ mass of 1.5 (1.0) TeV.
(a) the total lepton pair production cross section, (b) the forward-backward
asymmetry, (c) the left-right asymmetry, and (d) the polarized
forward-backward asymmetry.

\vskip 1.5truecm

\noindent
FIG. 3.

\noindent
Same as Fig.~2, but for $b$-quark production at the NLC.

\vfil\eject

\bye